\documentclass{gretsi}
\usepackage[english,french]{babel}   
\usepackage[T1]{fontenc}
\usepackage{amsmath,graphicx}
\usepackage{amssymb,amsfonts}
\usepackage{tikz}
\usepackage[mathscr]{eucal}
\usepackage{todonotes}

\usepackage[ruled,vlined]{algorithm2e}
\usepackage{multirow}
\usepackage{wrapfig}

\usepackage{xcolor}

\DeclareMathOperator*{\argmin}{arg\,min}

\title{Nonnegative Tucker Decomposition with Beta-divergence for Music Structure Analysis of Audio Signals}

\author{\coord{Axel}{Marmoret}{1},
        \coord{Florian}{Voorwinden}{1},
        \coord{Valentin}{Leplat}{2},
        \coord{Jérémy E.}{Cohen}{3},
        \coord{Frederic}{Bimbot}{1}}

\address{\affil{1}{Univ. Rennes 1, Inria, CNRS, IRISA, France.}
\affil{2}{Center for Artificial Intelligence Technology (CAIT), Skoltech, Moscow, Russia.}
\affil{3}{Univ Lyon, INSA-Lyon, UCBL, UJM-Saint Etienne, CNRS, Inserm, CREATIS UMR 5220, Villeurbanne, France}
}

\email{axel.marmoret@irisa.fr, jeremy.cohen@cnrs.fr}

\newcommand{\drawmatr}[6]{
    \draw[black,fill=#6] (#1,#2,#3) -- ++(#4,0,0) -- ++(0,-#5,0) -- ++(-#4,0,0) -- cycle;
}
\newcommand{\drawtens}[7]{
    \draw[black,fill=#7] (#1,#2,#3) -- ++(#4,0,0) -- ++(0,-#5,0) -- ++(-#4,0,0) -- cycle;
    \draw[black,fill=#7] (#1+#4,#2,#3) -- ++(0,0,-#6) -- ++(0,-#5,0) -- ++(0,0,#6) -- cycle;

    \draw[black,fill=#7] (#1,#2,#3) -- ++(#4,0,0) -- ++(0,0,-#6) -- ++(-#4,0,0) -- cycle;
}

\frenchabstract{La décomposition en Tucker nonnégatif (NTD), un modèle de décomposition tensorielle, permet d'extraire des motifs pertinents de manière non-supervisée, notamment dans le cadre de signaux audios. Néanmoins, les algorithmes actuels pour calculer la NTD sont souvent conçus pour optimiser la norme euclidienne. Ce travail propose un algorithme à mises à jour multiplicatives pour optimiser la NTD par rapport à la $\beta$-divergence, souvent considérée comme plus pertinente pour traiter des signaux audios. En particulier, cet article montre comment implémenter ces mises à jour de manière efficace en utilisant l'algèbre tensoriel. Finalement, des résultats expérimentaux sur la tâche d'estimation de la structure musicale montrent que la NTD optimisée par rapport à la $\beta$-divergence améliore les précédents résultats obtenus par rapport à la norme euclidienne.}

\englishabstract{Nonnegative Tucker decomposition (NTD), a tensor decomposition model, has received increased interest in the recent years because of its ability to blindly extract meaningful patterns, in particular in music information retrieval. Nevertheless, existing algorithms to compute NTD are mostly designed for the Euclidean loss. This work proposes a multiplicative updates algorithm to compute NTD with the $\beta$-divergence loss, often considered a better loss for audio processing. We notably show how to implement efficiently the multiplicative rules using tensor algebra. Finally, we show on a music structure analysis task that unsupervised NTD fitted with $\beta$-divergence loss outperforms earlier results obtained with the Euclidean loss.}

\begin{document}
%
\maketitle
\section{Introduction}
\label{sec:intro}
Tensor factorization models are powerful tools to interpret multi-way data, and are nowadays used in numerous applications~\cite{KoldaBader}. 
These models allow to extract interpretable information from the input data, 
generally in an unsupervised (or weakly-supervised) fashion, which can be a great asset when training data is scarcely available.
This is 
the case for music structure analysis (MSA) which consists in segmenting music recordings from the audio signal. For such applications, annotations can be ambiguous and difficult to collect~\cite{nieto2020segmentationreview}.

Nonnegative Tucker decomposition (NTD) has previously proven to be a powerful tool for MSA ~\cite{marmoret2020uncovering, smith2018nonnegative}.
While usually the Euclidean distance is used to fit the NTD, audio spectra exhibit large dynamics with respect to frequencies, which leads to a preponderance of few and typically low frequencies when using Euclidean distance. Contrarily, $\beta$-divergences, and more particularly Kullback-Leibler and Itakura-Saito divergences, are known to be better suited for time-frequency features. We introduce a new algorithm for NTD where the objective cost is the minimization of the $\beta$-divergence, and we study the resulting decompositions as regards their benefit on the MSA task on the RWC-Pop database~\cite{goto2002rwc}. The proposed algorithm adapts the multiplicative updates framework well-known for nonnegative matrix factorization (NMF)~\cite{fevotte2009nonnegative, lee1999learning} 
to the tensor case, detailing efficient tensor contractions. 
It is closely related to~\cite{kim2008nonnegative}, 
but studies instead the $\beta$-divergence and proposes modified multiplicative updates that guarantees global convergence to a stationary point. Code is fully open-source in \textit{nn\_fac}\footnote{https://gitlab.inria.fr/amarmore/nonnegative-factorization}.





\section{Mathematical background}
\label{sec:mathematical_aspects}
\subsection{Nonnegative Tucker Decomposition}
\begin{figure}[htbp]
\vspace{-10pt}
\begin{tikzpicture}[scale=0.8]
    \drawtens{0}{0}{0}{2}{2}{1.4}{white}
    \node at (2.8,-1) {$=$};
    \node at (1,-1,0){$\mathscr{X}$};
    \draw[->,black] (-0.15,-1.9,0)--(-0.15,-0.1,0);
    \node at (-0.4,-1,0) {\tiny $J$};
    \draw[->,black] (0.1,-2.15,0)--(1.9,-2.15,0);
    \node at (1,-2.35,0) {\tiny $K$};
    \draw[->,black] (-0.1,0.1,-0.1)--(-0.1,0.1,-1.3);
    \node at (-0.2,0.3,-0.4) {\tiny $L$};

    \drawmatr{3.4}{0}{0}{1.2}{2}{white};
    \draw[->,black] (3.3,-1.9,0)--(3.3,-0.1,0);
    \node at (3.1,-1,0) {\tiny $J$};
    \draw[->,black] (3.5,-2.1,0)--(4.5,-2.1,0);
    \node at (4,-2.3,0) {\tiny $J'$};

    \draw[black,fill=white] (5.6,0,0) -- ++(1,0,0) -- ++(0,0,-1.2) --
    ++(-1,0,0) -- cycle;
    \draw[->,black] (5.55,0.1,-0.1)--(5.55,0.1,-0.9);
    \node at (5.45,0.25,-0.4) {\tiny $L'$};
    \draw[->,black] (6.2,0.57,0)--(7,0.57,0);
    \node at (6.6,0.72,0) {\tiny $L$};

    \drawtens{5.2}{-0.5}{0}{1}{1}{0.9}{white}
    \draw[->,black] (5.1,-1.4,0)--(5.1,-0.6,0);
    \node at (4.9,-1,0) {\tiny $J'$};
    \draw[->,black] (5.3,-1.6,0)--(6.1,-1.6,0);
    \node at (5.7,-1.8,0) {\tiny $K'$};
    \draw[->,black] (5.05,-0.5,-0.1)--(5.05,-0.5,-0.9);
    \node at (5.1,-0.2,0) {\tiny $L'$};

    \drawmatr{7.1}{-0.5}{0}{2}{1}{white};
    \draw[->,black] (7,-1.4,0)--(7,-0.6,0);
    \node at (6.8,-1,0) {\tiny $K'$};
    \draw[->,black] (7.2,-1.6,0)--(9,-1.6,0);
    \node at (8.05,-1.8,0) {\tiny $K$};

    \node at (4,-1) {$W$};
    \node at (6.3, 0.2) {$Q^T$};
    \node at (8.05,-1) {$H^T$};
    \node at (5.7,-1) {$\mathscr{G}$};
\end{tikzpicture}
\vspace{-10pt}

\caption{Nonnegative Tucker decomposition of tensor $\mathscr{X}$ in factor matrices $W,H,Q$, and core tensor $\mathscr{G}$.}
\label{fig:NTD_decomp_tikz}
\end{figure}
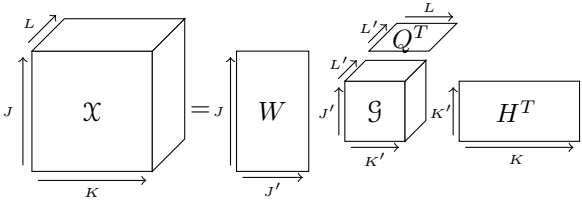

NTD is a mathematical model where a nonnegative tensor is approximated as the product of factors (one for each mode of the tensor) and a small core tensor linking these factors.
NTD is often used as a dimensionality reduction technique, but it may also be seen as a part-based representation similar to NMF.
In this work, we focus on third-order tensors for simplicity.
Denoting $\mathscr{X} \in \mathbb{R}_{+}^{J \times K \times L}$ the tensor to approximate and using conventional tensor-product notation~\cite{KoldaBader}, computing the NTD boils down to seek for three nonnegative matrices $W \in \mathbb{R}_{+}^{J \times J'}$, $H \in \mathbb{R}_{+}^{K \times K'}$ and $Q \in \mathbb{R}_{+}^{L \times L'}$ and a core tensor $\mathscr{G} \in \mathbb{R}_{+}^{J' \times K' \times L'}$ such that:
\begin{equation}
\label{NTD}
\begin{aligned}
\mathscr{X} &\approx \mathscr{G} \times_1 W \times_2 H \times_3 Q
\end{aligned}~
\end{equation}
This decomposition is also presented in Figure~\ref{fig:NTD_decomp_tikz}.

NTD is generally performed by minimizing some distance or divergence function between the original tensor and the approximation. Many algorithms found in the literature~\cite{marmoret2020uncovering, zhou2015efficient,phan2011extended} are based on the minimization of the squared Euclidean distance. In this work, we instead consider the $\beta$-divergence, detailed hereafter.

\subsection{The $\beta$-divergence loss function}


In this work, we will focus on the $\beta$-divergence function introduced in~\cite{Basu}. Given two nonnegative scalars $x$ and $y$, the $\beta$-divergence between $x$ and $y$ denoted $d_{\beta}(x|y)$ is defined as follows:
\begin{equation}
\begin{aligned}
d_{\beta}(x|y)=\left\{
\begin{array}{crl}
\frac{x}{y}-\log(\frac{x}{y})-1& \beta=&0\\
x\log(\frac{x}{y})+(y-x)&\beta=&1\\
\frac{x^{\beta}+(\beta-1)y^{\beta}-\beta xy^{\beta-1}}{\beta(\beta-1)} &\beta\in&\mathbb{R}\backslash\{0,1\}
\end{array}
\right.
\end{aligned}
\end{equation}
This divergence generalizes the Euclidean distance ($\beta=2$), and the Kullback-Leibler (KL) ($\beta=1$) and Itakura-Saito (IS) ($\beta=0$) divergences. The $\beta$-divergence $d_{\beta}(x|y)$ is homogeneous of degree $\beta$, that is for any $\lambda\in\mathbb{R}$, we have $d_{\beta}(\lambda x|\lambda y)=\lambda^{\beta}d_{\beta}(x|y)$.
It implies that factorizations obtained with $\beta > 0$ (such as the Euclidean distance or the KL divergence) will rely more heavily on the largest data values and less precision is to be expected in the estimation of the low-power components. The IS divergence ($\beta=0$) is scale-invariant and is the only one in the $\beta$-divergences family to possess this property. It implies that entries of low power are as important in the divergence computation as the areas of high power. This property is interesting when processing audio signals as low-power frequency bands can contribute as much as high-power frequency bands to their characterization.
Both KL and IS divergences are notoriously known to be better suited to audio source separation than the Euclidean distance~\cite{fevotte2009nonnegative,Fevotte_betadiv}.

Hence, this work focuses on how to compute a candidate solution to approximate NTD with $\beta$-divergence as a loss function:
\vspace{-5pt}
\begin{equation}\label{betaNTD_optiProb}
\underset{W\geq 0,H\geq0,Q\geq0,\mathscr{G}\geq 0}{\argmin} D_{\beta}(\mathscr{X}|\mathscr{G} \times_1 W \times_2 H \times_3 Q)
\end{equation}
with $D_{\beta}$ the elementwise $\beta$-divergence.

\section{A multiplicative updates algorithm}
\label{sec:algorithm}

\subsection{Multiplicative updates rules}
The cost function is non-convex with respect to all factors, and computing a global solution to NTD is NP-Hard since NTD is a generalization of NMF. 
However, each subproblem obtained when fixing all but one mode is convex as long as \(\beta\in[1,2]\). 
Hence, block-coordinate algorithms, that update one factor at a time while fixing all the other factors, are standard to solve both NMF and NTD~\cite{lee1999learning, phan2011extended, Fevotte_betadiv}. 
In particular, the seminal paper by Lee and Seung~\cite{lee1999learning} proposed an alternating algorithm for NMF with $\beta$-divergence, later revisited by Fevotte \textit{et. al.}~\cite{Fevotte_betadiv}, which we shall extend to NTD.

The multiplicative updates (MU) rule in approximate NMF \(M\approx UV^{\intercal}\) are 
\begin{equation}
    U \leftarrow \max\left( U\cdot\left(\frac{\left[(UV)^{\cdot(\beta-2)}\cdot M\right]V^{\intercal}}{(UV)^{\cdot(\beta-1)}V^{\intercal}}\right)^{.\gamma(\beta)}, \epsilon \right)~
\label{eq:nmf_mu}
\end{equation}
with $\cdot$ and $\div$ the element-wise product and division, $\epsilon > 0$ a small constant and $\gamma(\beta)$ a function equal to $\frac{1}{2 - \beta}$ if $\beta < 1$, 1 if $1 \leq \beta \leq 2$, and $\frac{1}{\beta - 1}$ if $2 < \beta$~\cite{Fevotte_betadiv}. The element-wise maximum between the matrix update, \textit{i.e.} the closed form expression of the minimizer of the majorization built at the current iterate, and $\epsilon$ in \eqref{eq:nmf_mu} aims at avoiding zero entries in factors, which may cause division by zero, and establishing convergence guarantee to stationary points within the BSUM framework~\cite{doi:10.1137/120891009}.

\subsection{Multiplicative updates for NTD}

The NTD model can be rewritten using tensor matricization, \textit{e.g.} along the first mode:
\begin{equation}
\begin{aligned}
 \mathscr{X} &= \mathscr{G} \times_1 W \times_2 H \times_3 Q \\
 \Leftrightarrow\mathscr{X}_{(1)} &= W\mathscr{G}_{(1)}\left(H\otimes Q\right)^{\intercal}
\end{aligned}
\label{eq:unfolded}
\end{equation}
where $\mathscr{X}_{(i)}$ is the matricization of the tensor $\mathscr{X}$ along mode $i$~\cite{KoldaBader} and \(\otimes\) denotes the Kronecker product. The matricization are analogous for factors $H$ and $Q$. One can therefore interpret equation~\eqref{eq:unfolded} as a NMF of \(\mathscr{X}_{(1)}\) with respect to \(W\) and \(\mathscr{G}_{(1)}\left(H\otimes Q\right)\)$^{\intercal}$.

A difficulty is that forming the Kronecker products is bound to be extremely inefficient both in terms of memory allocation and computation time. 
Instead, for the MU rules of factor matrices $W,H,Q$, 
matrix $V:=\mathscr{G}_{(i)}\left(H\otimes Q\right)^{\intercal}$ can be computed efficiently using the identity:
\begin{equation}
\mathscr{G}_{(1)}\left(H\otimes Q\right)^{\intercal} 
    =(\mathscr{G} \times_2 H \times_3 Q)_{(1)}
\end{equation}
which brings down the complexity of forming $V$ from\footnote{The multiway products are computed in lexicographic order.} $\mathcal{O}(KLJ'K'L')$ if done naively to $\mathcal{O}(KJ'K'L'+LJ'KL')$ and drastically reduces memory requirements.


For the core factor, one can use the vectorization property
\begin{equation}
\mathrm{vec}(\mathscr{X}) = (W\otimes H\otimes Q)\mathrm{vec}(\mathscr{G})~,
\end{equation}
to relate the core update with the NMF MU rules.
Again matrix $U:=W\otimes H\otimes Q$ 
is $J'K'L'$ times larger than the data itself. Therefore, 
for any vector $t:=\mathrm{vec}(\mathscr{T})$ 
we use the identity
\begin{equation}
    \left(W\otimes H\otimes Q\right)t = \mathrm{vec}(\mathscr{T} \times_1 W\times_2 H\times_3 Q)~.
    \label{eq:core_update_product}
\end{equation}
Products $U^{\intercal}t$ are computed similarly.

Algorithm~\ref{alg:algo_ntd_mu} shows one loop of the proposed MU algorithm. 
The overall complexity of such an iteration is dependant on the multiway product effective complexity, but is no worse than \(\mathcal{O}(JKLJ')\) if \(J,K,L > J'>K', L'\).
The proposed MU rules do not increase the cost at each iteration 
and for any initial factors, every limit point 
is a stationary point \cite[Theorem 8.9]{doi:10.1137/1.9781611976410}. 

\vspace{-6pt}
\begin{algorithm}
\SetAlgoLined

 \caption{A loop of $\beta$\_NTD($\mathscr{X}$,dimensions,$\beta$)}
\KwIn{$\mathscr{X},\mathscr{G},W,H,Q,\epsilon,\beta$}
\KwOut{$\mathscr{G},W,H,Q$}

$V = (\mathscr{G} \times_2 H \times_3 Q)_{(1)}$

$W \leftarrow \max \left( W\cdot \left(\frac{\left[(WV)^{\cdot(\beta-2)}\cdot \mathscr{X}_{(1)}\right]V^{\intercal}}{(WV)^{\cdot(\beta-1)}V^{\intercal}}\right)^{.\gamma(\beta)}, \epsilon \right)$




Perform analogous updates for $H$ and $Q$

$\mathscr{N} = (\mathscr{G}\times_1 W\times_2 H\times_3 Q)^{\cdot(\beta-2)}\cdot \mathscr{X}$

$\mathscr{D} = (\mathscr{G}\times_1 W\times_2 H\times_3 Q)^{\cdot(\beta-1)}$

$\mathscr{G} \leftarrow \max \left(\mathscr{G}\cdot \left(\frac{\mathscr{N}\times_1 W^{\intercal}\times_2 H^{\intercal}\times_3 Q^{\intercal}}{\mathscr{D}\times_1 W^{\intercal}\times_2 H^{\intercal}\times_3 Q^{\intercal}}\right)^{.\gamma(\beta)}, \epsilon \right)$
\label{alg:algo_ntd_mu}
\end{algorithm}
\vspace{-20pt}


\section{Experimental Framework}
\begin{table*}[ht]
\caption{Segmentation results on the RWC Pop dataset~\cite{goto2002rwc}, with different loss functions. P, R and F respectively represent Precision, Recall and F-measure, based on the evaluation of correct and incorrect boundaries (in time). These metrics are computed with two tolerances for considering a boundary correct: 0.5s and 3s. These values are standard in MSA~\cite{nieto2020segmentationreview}. Core dimensions $J', K', L'$ are fitted among values \{8, 16, 24, 32, 40\} with two-fold cross-validation, by splitting the RWC-Pop dataset between even and odd songs.}
\begin{center}
\begin{tabular}{|l|l|l|l|l|l|l|l|l|}

\hline
Representation         & \multicolumn{2}{l|}{Technique}        & $P_{0.5}$ & $R_{0.5}$ & $F_{0.5}$ & $P_3$  & $R_3$  & $F_3$  \\ \hline \hline
Chromas (initial work~\cite{marmoret2020uncovering}) & \multicolumn{2}{l|}{HALS-NTD} & 55.3\% & 59.3\% & 56.6\% & 70.3\% & 75.1\% & 71.9\%  \\ \hline \hline
\multirow{3}{*}{Mel-spectrogram} & \multicolumn{2}{l|}{HALS-NTD ($\beta = 2$)}         & 45.9\% & 49.5\% & 47.2\%  & 68.1\% & 73.0\% & 69.7\% \\ \cline{2-9}
                       & \multirow{2}{*}{MU-NTD} &  $\beta = 1$ &  53.3\%  & 56.9\%  & 54.6\%  & 70.5\%  & 75.3\% & 72.2\%  \\ \cline{3-9} 
                       &                         & $\beta = 0$ & 51.1\% & 56.8\% & 53.3\%  & 69.9\% & 78.0\% & 73.1\% \\ \hline \hline
\multirow{3}{*}{NNLMS} & \multicolumn{2}{l|}{HALS-NTD ($\beta = 2$)}         & 50.5\% & 52.7\% & 51.1\%  & 71.2\% & 74.6\% & 72.2\% \\ \cline{2-9} 
                       & \multirow{2}{*}{MU-NTD} & $\beta = 1$ & \textbf{57.8\%} & \textbf{61.9\%} & \textbf{59.3\%} & 73.9\% & 79.3\% & 75.9\% \\ \cline{3-9} 
                       &                         & $\beta = 0$ & 55.9\% & 61.7\% & 58.1\% & \textbf{74.0\%} & \textbf{81.9\%} & \textbf{77.1\%} \\ \hline
\end{tabular}
\end{center}
\label{table:seg_results}
\vspace{-16pt}

\end{table*}
\subsection{NTD for music processing}

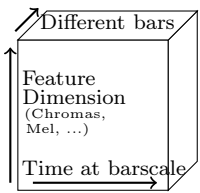
\begin{wrapfigure}{r}{0.35\columnwidth}
 \centering
  \begin{tikzpicture}
    \drawtens{0}{0}{0}{2}{2}{1}{white}
    \draw[->,thick,black] (-0.1,-1.9,0)--(-0.1,-0.1,0);
    \node at (0.55,-0.5,0) {\footnotesize{Feature}};
    \node at (0.75,-0.75,0) {\footnotesize{Dimension}};
    \node at (0.63,-1,0) {\tiny{(Chromas,}};
    \node at (0.53,-1.2,0) {\tiny{Mel, ...)}};
    \draw[->,thick,black] (0.2,-1.9,0)--(1.85,-1.9,0);
    \node at (1.15,-1.7,0) {\footnotesize{Time at barscale}};
    \draw[->,thick,black] (-0.15,0,-0.3)--(-0.15,0,-1.1);
    \node at (1,0.05,-0.5) {\footnotesize{Different bars}};
  \end{tikzpicture}
  \caption{TFB tensor}
  \vspace{-5pt}
\end{wrapfigure}
NTD has already been introduced to process audio signals, and was shown to provide a barwise pattern representation of a music~\cite{marmoret2020uncovering, smith2018nonnegative}. 
For music, NTD is performed on a 3rd-order tensor, called TFB tensor, which is the result of splitting a spectrogram on bar frontiers and concatenating the subsequent barwise spectrograms on a 3rd-mode. Hence, a TFB tensor is composed of a frequential mode and two time-related modes: an inner-bar (low-level) time, and a bar (high-level) time. 
Each bar contains 96 frames, which are selected as equally spaced on a oversampled spectrogram (hop length of 32 samples), in order to account for bar length discrepancies~\cite{marmoret2020uncovering}.

In our previous work~\cite{marmoret2020uncovering}, we computed the NTD on chromagrams, then evaluated on music segmentation. 
We extend this previous study to Mel-spectrograms
, which consist in the STFT of a song with frequencies aggregated following a Mel filter-bank. They provide a richer representation of music than chromagrams but they live in a higher dimensional space. On the basis of this alternate representation, we compare the algorithm introduced in~\cite{marmoret2020uncovering} (HALS-based NTD with Euclidean loss minimization, see~\cite{phan2011extended}) with the proposed algorithm in the MSA task, on the audio signals of the RWC Pop database~\cite{goto2002rwc}, which is a standard dataset in music information retrieval.

In practice, Mel-spectrograms 
are dimensioned following the work of~\cite{grill2015music}, which is considered as state-of-the-art in this task. Precisely, STFT are computed as power spectrograms with a window size of 2048 samples for a signal sampling rate of 44.1~kHz. A Mel-filter bank of 80 triangular filter between 80~Hz and 16~kHz is then applied. In addition to this raw Mel representation, we study a logarithmic variant, which is generally used as a way to account for the exponential distribution of power in audio spectra. As the logarithmic function is negative for values lower than 1, we introduce the Nonnegative Log-Mel spectrogram (NNLMS) as NNLMS = $\log(\mathrm{Mel} + 1)$. 

Finally, each nonnegative TFB tensor has sizes 80$\times$96$\times L$, with $L$ the number of bars in the song. When (empirically) setting core dimensions $J', K', L' = 32$, one iteration of the MU algorithm takes approximately 0.2s, while one iteration of the HALS algorithm takes approximately 0.75s on an Intel\textregistered  Core\texttrademark i7 processor. 
Nonetheless, the HALS algorithm generally converges in fewer iterations than MU.
\vspace{-13pt}
\subsection{Music structure analysis based on NTD}
\label{sec:structural_segmentation}
MSA consists in segmenting a song into sections (such as “verse”, “chorus”, etc) as presented in~\cite{nieto2020segmentationreview}. The goal here is to retrieve the boundaries between different sections. We use the same segmentation framework than in~\cite{marmoret2020uncovering}.

Segmentation results are presented in Table~\ref{table:seg_results}, where we compare the performance of segmenting the chromagram using the HALS-NTD~\cite{marmoret2020uncovering} with segmenting Mel and NNLMS representations. These results show that, for both representations, using the KL and IS divergences instead of the Euclidean loss enhance segmentation performance. Segmentation results are also higher when using NTD on the NNLMS rather than on the Mel-spectrogram, and outreach previous results on Chromagrams. Hence, adapting the decomposition to the dynamics of audio signals seems beneficial, both in term of loss function and feature.

\subsection{Qualitative assessment of patterns}
\label{sec:audio_patterns}
As a qualitative study between the different $\beta$-divergences ($\beta \in \{0,1,2\}$), we computed the NTD with these three values on the STFT of the song ``Come Together'' by the Beatles. Using the Griffin-Lim algorithm~\cite{griffin1984signal} and softmasking~\cite{smith2018nonnegative}, 
spectrograms computed with the NTD (such as musical patterns $W \mathscr{G}_{[:,:,i]} H^T$~\cite{marmoret2020uncovering}) 
are reconstructed into listenable signals. Results are available online\footnote{https://ax-le.github.io/resources/examples/ListeningNTD.html}, and qualitatively confirm that the KL divergence is better adapted to signals than the Euclidean loss, while this is more contrasted for the IS divergence.

\section{Conclusion}
\label{sec:conclusion}
Nonnegative Tucker decomposition is able to extract salient patterns in numerical data. This article has proposed a tractable and globally convergent algorithm to perform the NTD with the $\beta$-divergence as loss function. 
This appears to be of particular interest for a wide range of signals and applications, notably in the audio domain, as supported in this paper by quantitative results on a music structure analysis task and qualitative examples for $\beta = 1,0$.

Future work may consider the introduction of sparsity constraints, which generally improve the interpretability of nonnegative decompositions, and seeking additional strategies to accelerate the algorithm itself.

\small
\bibliographystyle{IEEEbib}
\bibliography{biblio}

\end{document}